# Unveiling van Hove singularity modulation and fluctuated charge order in kagome superconductor CsV$_3$Sb$_5$ via time-resolved ARPES

Yigui Zhong[1,*], Takeshi Suzuki[1], Hongxiong Liu[2], Kecheng Liu[1], Zhengwei Nie[2], Youguo Shi[2], Sheng Meng[2], Baiqing Lv[3], Hong Ding[3], Teruto Kanai[1], Jiro Itatni[1], Shik Shin[1], Kozo Okazaki[1,4,5,*]

[1]The Institute for Solid State Physics, The University of Tokyo, Chiba 277-8581, Japan

[2]Beijing National Laboratory for Condensed Matter Physics and Institute of Physics, Chinese Academy of Sciences, Beijing 100190, China

[3]Tsung-Dao Lee Institute and School of Physics and Astronomy, Shanghai Jiao Tong University, Shanghai 200240, China

[4]Trans-scale Quantum Science Institute, The University of Tokyo, Bunkyo, Tokyo 113-0033, Japan.

[5]Material Innovation Research Center, The University of Tokyo, Kashiwa, Chiba 277-8561, Japan.

Abstract

Kagome superconductor CsV$_3$Sb$_5$, which exhibits intertwined unconventional charge density wave (CDW) and superconductivity, has garnered significant attention recently. Despite extensive static studies, the nature of these exotic electronic orders remains elusive. In this study, we investigate the non-equilibrium electronic structure of CsV$_3$Sb$_5$ via time- and angle-resolved photoemission spectroscopy. Our results reveal that upon laser excitation, the van Hove singularities immediately shift towards the Fermi level and subsequently oscillate in sync with a 1.3 THz coherent phonon mode. By analyzing the coherent intensity oscillations in the energy-momentum ($E$-$k$) map, we find that this coherent phonon is strongly coupled with electronic bands from both Sb and V orbitals. While typically observable only in the CDW state, remarkably, we find that the 1.3-THz coherent phonon mode can be persistently excited at temperatures above $T_{CDW}$, suggesting the potential existence of fluctuated CDW in CsV$_3$Sb$_5$. These findings enhance our understanding of the unconventional CDW in CsV$_3$Sb$_5$ and provide insights into the ultrafast control of kagome superconductivity.

Elucidating the fundamental mechanisms behind the exotic electronic orders is a crucial issue in quantum materials research, as it not only deepens our knowledge of physics but also plays an important role in advancing material applications. The kagome lattice system, characterized by its unique geometry of corner-sharing triangles, features distinctive electronic structures as showing in Fig. 1 (a) such as a topological flat band, a Dirac band, and van Hove singularities (VHS) [1]. These electronic features make the kagome lattice a fertile platform for the emergence of exotic electronic orders [2-5]. Recently, significant attention has been drawn to the kagome metals $A$V$_3$Sb$_5$ ($A$ = K, Rb, Cs) [6-10], which exhibit a diverse array of electronic phenomena, including charge density wave (CDW) [11-14], superconductivity [15-19], nematic order [20,21], pair density wave [15], and the giant anomalous Hall effect [22,23]. Moreover, applying pressure or chemical doping to these kagome metals gradually suppresses the CDW while enhancing superconductivity, resulting in the appearance of two superconducting domes [16,17,24]. This highlights the intimate interplay between CDW and superconductivity, making a comprehensive understanding of the CDW crucial for elucidating the nature of superconductivity in these materials.

The origin of the CDW in $A$V$_3$Sb$_5$ is highly debated. Angle-resolved photoemission spectroscopy (ARPES) experiments have revealed the presence of the VHS near the Fermi level ($E_F$) with favorable nesting conditions [7,25-28], supporting a scenario driven by electronic instabilities. This is further corroborated by the observations of the time-reversal symmetry breaking [13,14] and the absence of the acoustic phonon anomaly [29] in the CDW state. In contrast, density functional theory (DFT) calculations [30,31] have identified two unstable phonon modes at M (1/2, 0, 0) and L (1/2, 0, 1/2) in Brillouin zone (BZ), indicating the potential lattice instabilities. Scanning tunneling microscopy [11,32,33] and x-ray diffraction measurements [7,29,34,35] have indeed observed 2×2×2 or 2×2×4 lattice distortions. Investigating the CDW in a static state poses challenges in determining whether it originates from the electron or lattice instabilities, as charge order and lattice distortions are intricately liked due to significant electron-phonon couplings. To address these challenges, time-resolved studies offer a promising alternative [36]. Ultrafast laser excitation can disturb or melt the CDW instantaneously [37,38], enabling the study of CDW recovery and the associated electron-phonon couplings. Furthermore, intense laser excitations can modulate the lattice and electron band structure in an ultrafast manner, possibly leading to the hidden quantum phenomena that are inaccessible under static conditions [39].

Recent time-resolved reflectivity studies [31,40] on CsV$_3$Sb$_5$ have demonstrated the emergence of a 1.3 THz coherent phonon mode below CDW transition temperature ($T_{CDW} \sim 94$ K) and a 3.1 THz coherent phonon mode at a lower temperature of approximately 60 K. These findings underscore the strong couplings between the emerging coherent phonons and electrons. However, detailed information regarding such couplings in energy and momentum spaces remain elusive. These details can be elucidated by performing time- and angle-resolved photoemission spectroscopy (TARPES) measurements [39,41], which directly capture the dynamics of electron band structure and investigate how the electronic bands are affected by coherent phonon vibrations. By mapping the Fourier component of TARPES intensity, known as frequency-domain ARPES [42,43], it is possible to obtain comprehensive insights about the band- and mode-resolved couplings.

In this letter, we study the band-resolved couplings between coherent phonons and electrons on a kagome superconductor CsV$_3$Sb$_5$ utilizing TARPES. Upon ultrafast pulse excitations, we first observe the hot electrons excited to the unoccupied states above $E_F$ and remarkable energy shifts of the electronic bands. Particularly, the VHS significantly shift upward to $E_F$ after absorbing laser pulse. Subsequently, the VHSs decay rapidly in 0.2 ps with a superimposed oscillation at a frequency of approximately 1.3 THz. This unambiguously suggests a strong coupling between the VHS and a 1.3 THz coherent phonon that is linked to the CDW as confirmed by static Raman and time-resolved reflectivity spectroscopy measurements [31,40,44]. By studying the temporal photoemission changes in different momenta and energies, we further map out the frequency-domain ARPES spectra and find this 1.3-THz phonon universally couples with the electronic bands originated from both Sb and V atoms. Remarkably, we find this coherent phonon mode is persistently observable after the long-range CDW order suppressed by increasing temperature to $T > T_{CDW}$, implying the potential existence of the fluctuated CDW order in kagome superconductor CsV$_3$Sb$_5$. Our findings complement to the static studies and are helpful in comprehensively understanding the nature of the unconventional CDW and exploring ultrafast modulation of kagome superconductivity.

The synthesis and characterization of high-quality crystal $CsV_3Sb_5$ was described in our previous work [19]. For TARPES measurements [45], a commercial Ti:Sapphire regenerative amplifier system (Spectra Physics, Solstice Ace) with a center wavelength of 800 nm (hν = 1.55 eV) and pulse duration of 35 fs was used for the pump light. The fluence of the pump laser is fixed at 0.7 mJ/cm$^2$ in this study except for a specific statement. After generating a second harmonic via 0.2-mm-thick β-BaB$_2$O$_4$, the light was focused into a gas cell filled with Ar and high harmonics were generated. The seventh harmonic light (hν = 21.7 eV) were used as the probe light. The repetition rate was set to 10 kHz. The temporal resolution was evaluated to be 55 fs from the TARPES intensity far above $E_F$ corresponding to the cross correlation between pump and probe pulses. The combined energy resolution was set to 150 meV for the TARPES measurements. The static ARPES measurements were performed using a Helium discharge lamp and the energy resolution was set to be 12.5 meV. Both in TARPES and static ARPES measurements, a Scienta R4000 hemispherical electron analyzer was used to collect photoelectrons.

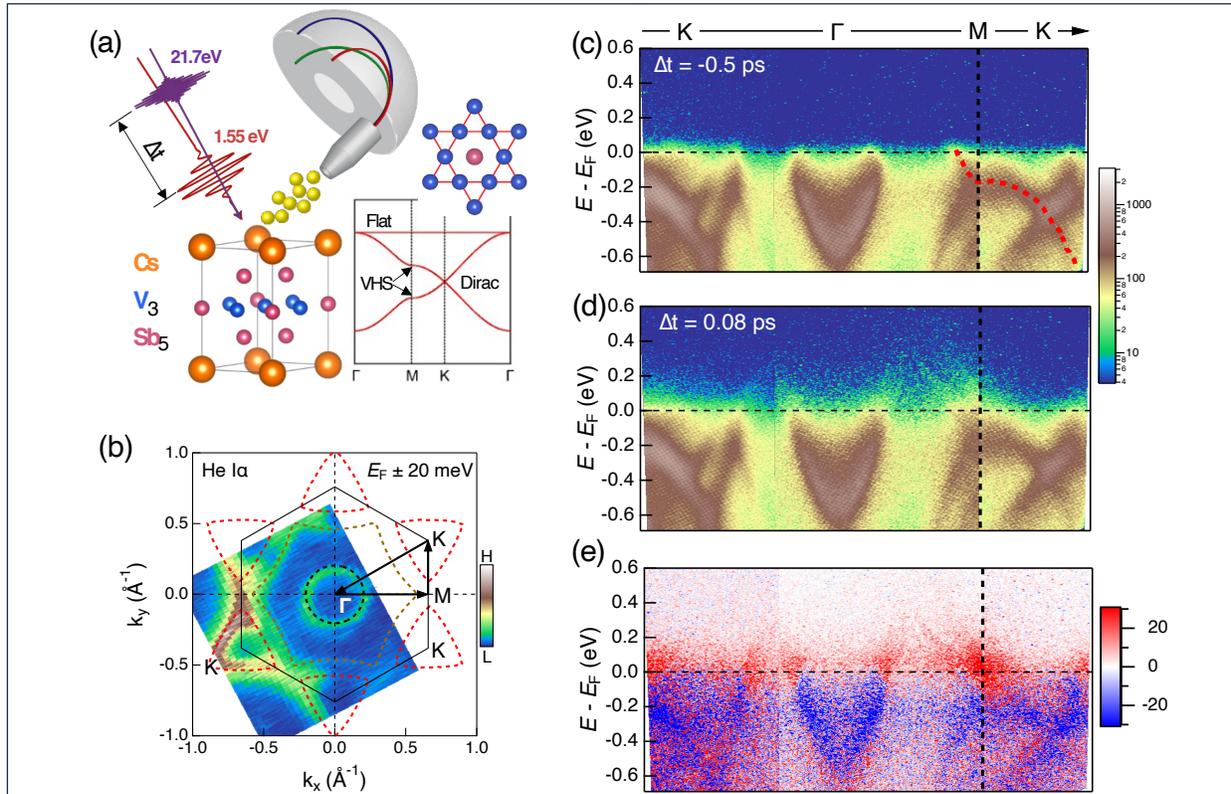

**Fig. 1.** (a) Schematic plots for electronic band structure of a kagome lattice, TARPES experimental setup, and the crystal structure of $CsV_3Sb_5$. (b) Fermi surface acquired by a static ARPES using Helium discharge lamp (hν = 21.218 eV). The dashed lines represent the contours of the Fermi surface. (c) ARPES spectra along with K-Γ-M-K cut showing as black arrows in (b) measured using high-hormonic generated ultrafast laser hν = 21.7 eV before the arrival of pump pulse. (d) Same with (c) but taken at 0.08 ps after pumped by 1.55-eV laser pulse. (e) The differential plot between ARPES intensities before and after the laser pump. The intensity in (c) and (d) is plotted in a logarithmic scale as the color bar shown beside. The dashed red line in (c) traces the saddle band at M point.

The crystal structure of $CsV_3Sb_5$ is shown in Fig. 1(a). It has a layer structure that consists of V-Sb layer in which V atoms form a kagome lattice, Sb layer, and Cs layer along $c$-axis in the unit cell of the normal state. Figure 1(b) shows the Fermi surface (FS) of $CsV_3Sb_5$ measured by static ARPES with a photon energy of 21.218 eV (He I$\alpha$). It includes three FS sheets occupied by electrons from V-Sb layer: a circular FS centered at $\Gamma$ that is derived from Sb-5$p_z$ orbital, and a large hexagonal FS and a small triangular FS that are derived from V-3$d$ orbitals. By taking a loop cut along K–$\Gamma$–M–K direction, Fig. 1(c) shows the overall electronic band structure in the equilibrium state. The electronic bands, including an electron-like parabolic band at $\Gamma$, a Dirac-like band at K and a VHS at M, are consistent with previous static ARPES measurements and DFT calculations [7,25-28,30]. We note that previously reported two VHSs near $E_F$ at M are indistinguishable here because of the limited energy resolution in TARPES measurements.

With establishing the consistency of the band structure in the equilibrium state, we now turn to study the non-equilibrium band structure at $T < T_{CDW}$ measured by TARPES [see a schematic plot in Fig. 1(a)]. At a delay time $\Delta t$ = 0.08 ps after the arrival of the pump pulse, it clearly tells from the band structure shown in Fig. 1(d) that the hot electrons are excited to occupy the states above $E_F$. To better visualize the induced changes in the photoemission intensity, the differential plot between the spectra before and after the arrival of the pump pulse is shown in Fig. 1(e). Near $E_F$, the intensity change is primarily due to the significantly elevated electronic temperature resulting from the absorption of the pump pulse. Notably, in addition to this electronic temperature effect, there are the remarkable energy shifts in the electronic bands. A clear increase of the intensity between $E_F$ and the saddle band at M indicates that the VHS shifts up toward $E_F$. Simultaneously, a clear increase in the intensity below the bottom of the parabolic band at $\Gamma$, along with the enlarged Fermi momentum, collectively suggest a downward shift of the parabolic band.

Given the importance of the energy of the VHS to physical properties of kagome materials, we closely examine the temporal evolution of the VHS at M. The extracted energy distribution curves (EDCs) at different delay times are presented as an image in Fig. 2(a). Clearly, the peak of the EDCs at M, representing the energy position of the VHS, shifts immediately toward to $E_F$ upon laser excitation. To quantitatively extract temporal evolution of this energy shift, we fitted the EDCs with a Lorentzian function multiplied by a Fermi-Dirac distribution (details shown in supplementary materials). From these fits, we determined the energy shift of the VHS as function of delay time, as shown in Fig. 2(b). The results exbibits an exponential decay with a superimposed oscillation. Just after the excitation, the VHS shifts upward by approximately 58 meV. This VHS shift was observed in static ARPES measurements [27,46] when temperature across $T_{CDW}$, thus, indicating that laser excitation effectively melts the CDW in our measurements.

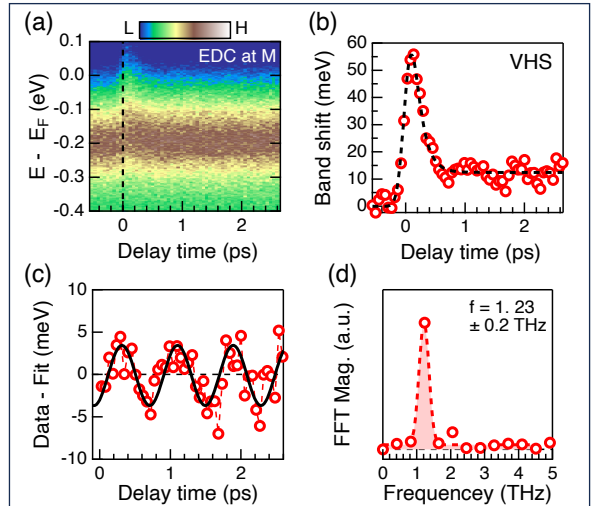

**Fig. 2.** (a) EDCs at M point as function of delay time. (b) Temporal energy shift of the van Hove singularity (VHS) extracted from the fits to EDCs at M point. The dashed line is an exponential decay fit. (c) Remained oscillation component after subtracting an exponential decay fit shown in (b). The black line is the fit to a damped cosine function. (d) Magnitude of the Fast Fourier transformation (FFT) on the oscillation component in (c).

By fitting the dynamics of the VHS shifts in Fig. 2(b) to a single-exponential decay function, we find the relaxation time of the VHS to be approximately 150 fs. After subtracting the decay background, the oscillation component is extracted, as shown in Fig. 2(c). Fast Fourier transformation (FFT) analysis of the oscillation, presented in Fig. 2(d), reveals a predominant frequency of 1.23 ± 0.2 THz. Considering the experimental uncertainties, this value is close to the frequency of the coherent phonon (~1.3 THz) observed in previous time-resolved reflectivity spectroscopy [31,40]. Our result directly demonstrates a strong coupling between the VHS and the 1.3-THz coherent phonon mode. Moreover, the VHS oscillation resembles a cosine function [Fig. 2(c)]. Combined with the immediately modified density of states (DOS) after laser excitation, manifested as elevated electronic temperature (see supplementary materials) and band shift, the excitation of this coherent phonon mode is most likely consistent with the mechanism of displacive excitation of coherent phonons (DECP) [47].

Utilizing the advantage of the momentum and energy resolution of TARPES, we next study the momentum dependence of coherent phonon excitations and their band-resolved couplings with electrons by analyzing the temporal photoemission intensity change. At $T$ = 8 K, well below $T_{CDW}$, the photoemission intensity changes in Γ, M, and K areas are integrated near $E_F$ (from - 0.025 eV to + 0.125 eV) over the momentum ranges shown by the solid red lines in Fig. 3(a). The intensity change is normalized to the intensity before the arrival of the pump ($I_0$) and is thus defined as $\Delta I/I_0(t)$. As shown in Fig. 3(b), the intensity oscillation can be observed in both Γ, M and K areas.

To extract the oscillation part, we first fit these intensity dynamics $\Delta I/I_0(t)$ with the exponential decay function convolved with a

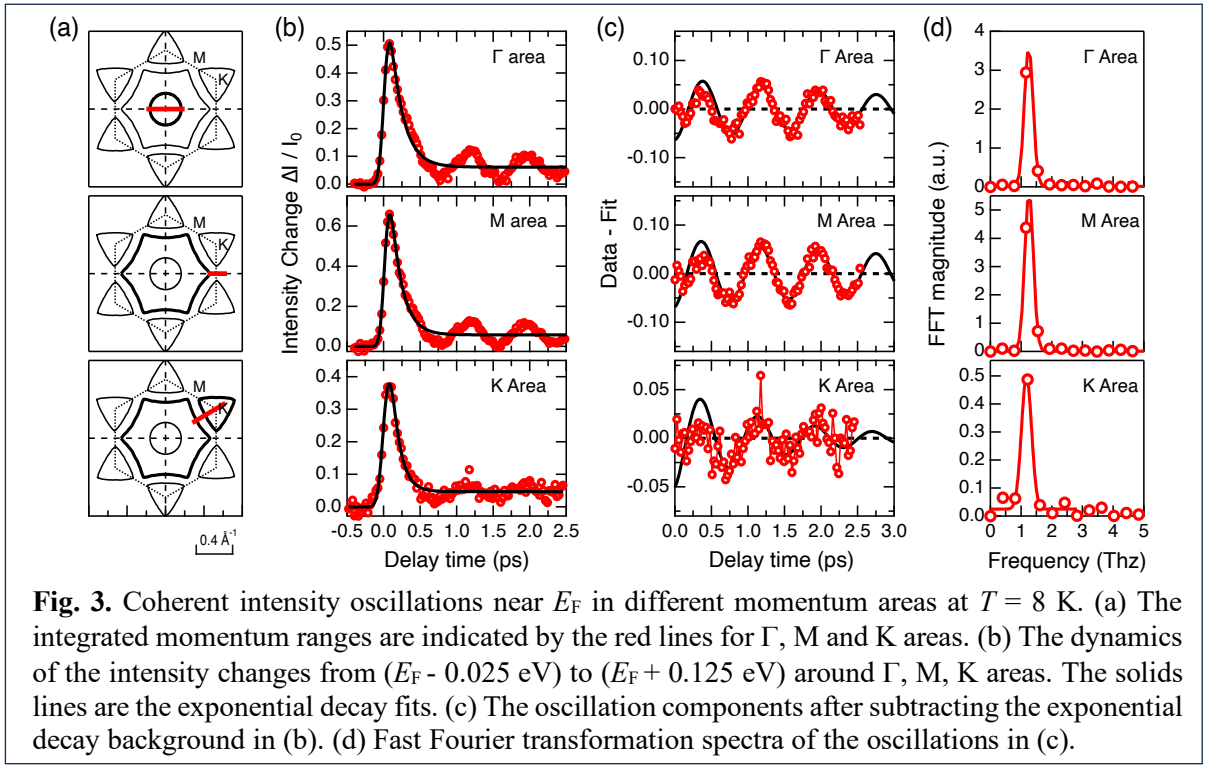

**Fig. 3.** Coherent intensity oscillations near $E_F$ in different momentum areas at $T$ = 8 K. (a) The integrated momentum ranges are indicated by the red lines for Γ, M and K areas. (b) The dynamics of the intensity changes from ($E_F$ - 0.025 eV) to ($E_F$ + 0.125 eV) around Γ, M, K areas. The solids lines are the exponential decay fits. (c) The oscillation components after subtracting the exponential decay background in (b). (d) Fast Fourier transformation spectra of the oscillations in (c).

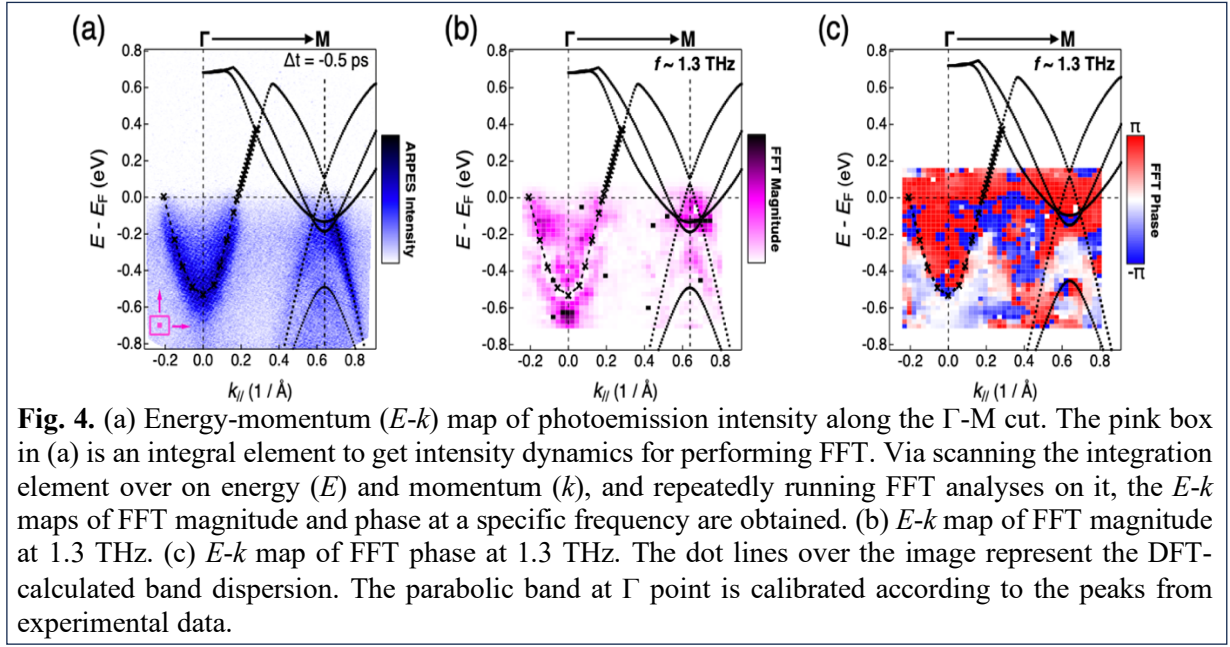

**Fig. 4.** (a) Energy-momentum (*E-k*) map of photoemission intensity along the Γ-M cut. The pink box in (a) is an integral element to get intensity dynamics for performing FFT. Via scanning the integration element over on energy (*E*) and momentum (*k*), and repeatedly running FFT analyses on it, the *E-k* maps of FFT magnitude and phase at a specific frequency are obtained. (b) *E-k* map of FFT magnitude at 1.3 THz. (c) *E-k* map of FFT phase at 1.3 THz. The dot lines over the image represent the DFT-calculated band dispersion. The parabolic band at Γ point is calibrated according to the peaks from experimental data.

Gaussian function representing time resolution. The best fits are shown as the black lines in Fig. 3(b). By subtracting the exponential decay background, the oscillation components are isolated and summarized in Fig. 3(c). FFT analyses of these oscillations yield the frequency-domain spectra shown in Fig. 3(d). The results indicate that in Γ and M areas a single frequency is observed with $f_{\Gamma,M} \sim 1.3 \pm 0.2$ THz with similar FFT magnitudes. The oscillation frequency in K areas is observed at $f_K \sim 1.2 \pm 0.2$ THz. The small difference of 0.1 THz compared to $f_{\Gamma,M}$ could be due to the signal sampling during FFT process. Thus, the coherent phonon with a frequency of 1.3 THz, which would be vibrations of Cs atoms as discussed later, is commonly observed near $E_F$ in Γ, M and K areas. This coherent phonon is consistent with the previous time-resolved optical measurements [31,40]. However, other coherent phonons with higher frequencies (~3.1 THz and ~4.1 THz) reported in time-resolved optical spectroscopy are absent, likely quenched by the relatively higher fluence used in this study (~0.7 mJ/cm$^2$). Moreover, the phase of the oscillation components shown in Fig. 3(c) is determined to be π through fitting to a damped cosine function, further supporting the DECP mechanism.

To obtain the energy-momentum (*E-k*) distributions of the coherent oscillation caused by the 1.3-THz phonon mode, we further map out the frequency-domain APRES spectra $I(E, k, f = 1.3 \text{ THz})$, which are obtained from the FFT analyses on TARPES spectra $I(E, k, t)$. In Figs. 4(b) and 4(c), we present the *E-k* distributions of the FFT magnitude and phase at $f = 1.3$ THz along the Γ-M cut. Clearly, the FFT-magnitude mapping in Fig. 4(b) closely resembles the non-perturbated band dispersion shown in Fig. 4(a), suggesting universal couplings between the band dispersions along Γ-M and the 1.3-THz coherent phonon mode. Additionally, the band dispersion algins well with the boundary where the phase of FFT changes from ±π (negative cosine) to 0 (positive cosine), indicating the bands integrally oscillate following this coherent phonon. Since the parabolic band at Γ is derived from Sb-5$p_z$ orbital and the bands at M are derived from V-3d orbitals [25,26], these results indicate the 1.3-THz coherent phonon mode, which relates to the CDW according to previous Raman and time-resolved optical studies [31,40,44], is strongly coupled with both the Sb-5$p_z$ orbital-derived and V-3d orbital-derived bands, highlighting the three-dimensional feature of the CDW order in CsV$_3$Sb$_5$.

It is interesting to examine the 1.3-THz phonon mode at $T > T_{CDW}$ as it relates to the CDW. In Fig. 5(a), we compare the intensity dynamics at $T = 8$ K ($< T_{CDW}$) and $T = 150$ K ($> T_{CDW}$). The intensities are integrated from $E_F$ to $E_F + 0.15$ eV along with the Γ-M direction as the phonon couplings are in similar extents along this direction [Fig. 3(d)]. By subtracting the exponential decay background, Fig. 5(b) shows the remained intensity oscillations. Surprisingly, the coherent oscillation is persistently observable even at $T = 150$ K, although the amplitude noticeably weakens. Further FFT analysis shown in Fig. 5(c) demonstrates the oscillation at $T = 150$ K has the same frequency as the one at $T = 8$ K, but the magnitude decreases by approximately 5 times.

According to the previous Raman and time-resolved optical spectroscopy measurements [31,40,44], the sudden disappearance of the 1.3-THz phonon at $T = T_{CDW}$ as temperature increases has been observed. Furthermore, its frequency matches the frequency of a Cs motion at zone corner according to DFT calculations [31] and becomes larger in $A$V$_3$Sb$_5$ compounds with lighter alkali elements (Rb and K) [44]. These behaviors together suggest this phonon is corresponding to a zone-folded phonon with involving Cs motion. Below $T_{CDW}$, it is folded to the Γ point (q = 0) due to the periodic lattice distortion (PLD). Therefore, it becomes detectable only when $T < T_{CDW}$ in Raman spectroscopy, as it detects phonon modes at q = 0, as shown in Fig. 5(d). Generally, the optically excited coherent phonon mode is also constrained at zero-momentum (q = 0) [48]. This explains why the 1.3-THz coherent phonon mode is only excitable at $T < T_{CDW}$ by near-infrared pulse employed in the time-resolved reflectivity measurements [31,40]. The consistent phonon frequency, as well as the same constraint in detectable phonon momentum, demonstrates that our TARPES detect the same 1.3-THz phonon mode as that in static Raman spectroscopy and time-resolved reflectivity spectroscopy. Therefore, the most plausible scenario to explain our TAPRES observation of the 1.3-THz coherent phonon at $T > T_{CDW}$ is the existence of the fluctuated CDW order, which gives rise to fluctuated PLD.

The fluctuated PLD contributes to the finite folded density of the 1.3-THz phonon mode at q=0, making it observable in our TARPES via its couplings to electrons. In fact, the X-ray scattering studies [35] demonstrate a weak intensity of the CDW superlattice peak at [1.5, 1.5, ± 0.5] above $T_{CDW}$, suggesting the existence of the fluctuated 2×2×2 PLD. However, it is not detectable in the Raman spectroscopy due to its insensitivity to a small amplitude of the structural modulation induced by fluctuated CDW. In the non-equilibrium case, according to the so-called DECP mechanism, upon absorbing a laser pulse, the electronic DOS will be redistributed, exerting a force to drive in-sync vibrations of the atoms following the symmetry of the strongly coupled phonon modes. Under this mechanism, the higher fluence injected, the larger DOS will be redistributed, resulting in a stronger force to drive atomic vibrations. The relatively large fluence used in our TARPES, 0.7 mJ/cm$^2$, could successfully amplify the coherent atomic vibrations corresponding to the 1.3-THz phonon mode in the fluctuated CDW state, thus making it observable via strong electron-phonon couplings.

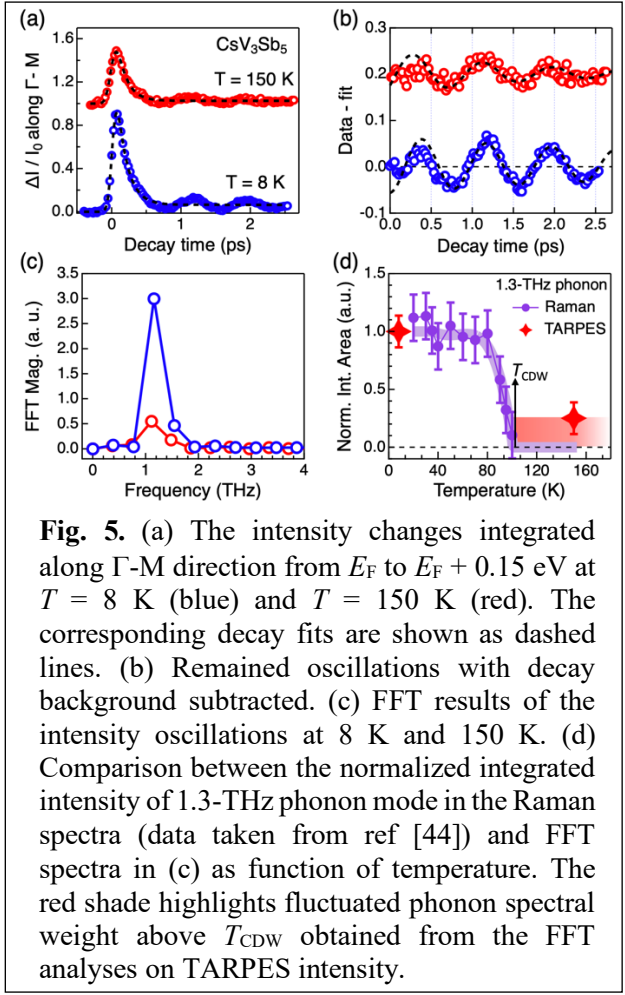

**Fig. 5.** (a) The intensity changes integrated along Γ-M direction from $E_F$ to $E_F + 0.15$ eV at $T = 8$ K (blue) and $T = 150$ K (red). The corresponding decay fits are shown as dashed lines. (b) Remained oscillations with decay background subtracted. (c) FFT results of the intensity oscillations at 8 K and 150 K. (d) Comparison between the normalized integrated intensity of 1.3-THz phonon mode in the Raman spectra (data taken from ref [44]) and FFT spectra in (c) as function of temperature. The red shade highlights fluctuated phonon spectral weight above $T_{CDW}$ obtained from the FFT analyses on TARPES intensity.

We note the 1.3-THz coherent phonon is not observed at $T > T_{\mathrm{CDW}}$ in time-resolved reflectivity studies [31,40], which utilized a much weak fluence (< 0.1 mJ/cm$^2$). The possible reason is the pump fluence does not exceed the threshold to excite the weak zone-folded coherent phonon in the fluctuated CDW state. Additionally, it is worth noting that previous time-resolved reflectivity studies detect the changes in dielectric constant in the near-infrared region, which may not be affected by excitation of 1.3-THz coherent phonon mode in the fluctuated CDW state, unlike TARPES, which is directly detects the electronic DOS and is much more sensitive to the change of DOS induced by the coherent excitations of a fluctuated zone-folded phonon.

In conclusion, our TAPRES studies on kagome superconductor CsV$_3$Sb$_5$ reveals a significant coupling between the electronic structure, particularly the VHS, and a 1.3-THz phonon mode associated with the CDW order. Notably, as the long-range CDW order is suppressed with increasing temperature beyond $T_{\mathrm{CDW}}$, our measurements reveal the persistent excitation of 1.3-THz coherent phonon, contrasting with previous time-resolved optical studies and suggesting the presence of the fluctuated CDW. Our findings have important implications for understanding the CDW order in kagome superconductors CsV$_3$Sb$_5$. The observed strong VHS-phonon coupling may result in a strong interplay between electron and lattice instabilities, thereby making the electron-phonon couplings and electron-electron interactions cooperatively generate the CDW with exotic properties. Furthermore, our results have important implications for the ultrafast optical modulation of kagome superconductors' physical properties. For instance, we demonstrate that the VHS can be ultrafast-tuned closer to $E_{\mathrm{F}}$, which increases the DOS near $E_{\mathrm{F}}$ and potentially leads to the light-enhanced superconductivity. Additionally, as the coherent phonon remains accessible through the DECP mechanism above $T_{\mathrm{CDW}}$, applying a stronger laser fluence could further perturb the lattice structure and enable the realization of the new quantum states unattainable under equilibrium conditions.

We thank Xun Shi and Shangfei Wu for fruitful discussions. This work at the University of Tokyo was supported by Grants-in-Aid for Scientific Research (KAKENHI) (Grant Nos. JP19H01818, JP19H00659, JP19H00651, JP24K01375, JP24K00565, and JP24KF0021) from the Japan Society for the Promotion of Science (JSPS), by JSPS KAKENHI on Innovative Areas "Quantum Liquid Crystals" (Grant No. JP19H05826), and the Quantum Leap Flagship Program (Q-LEAP) (Grant No. JPMXS0118068681) from the Ministry of Education, Culture, Sports, Science, and Technology (MEXT). This work at the Chinese Academy of Sciences was supported by the National Natural Science Foundation of China (Grants No. U22A6005, No. U2032204), the Strategic Priority Research Program of the Chinese Academy of Sciences (Grants No. XDB33010000), and the Synergetic Extreme Condition User Facility (SECUF). Y. Zhong is supported by JSPS International Researcher Fellowship program.

*Corresponding authors:

yigui-zhong@issp.u-tokyo.ac.jp

okazaki@issp.u-tokyo.ac.jp